\documentclass{jsara}
\journalinfo{Journal of the Southeastern Association for Research in Astronomy,
{\bf 7}, 51-55, December 1}

\slugcomment{}

\shorttitle{SX Phe in M107}
\shortauthors{McCombs et al.}

\begin{document}

%% This will be set in the editorial office to the starting number of your paper in the journal
\setcounter{page}{51}

%% LaTeX will automatically break titles if they run longer than
%% one line. However, you may use \\ to force a line break if
%% you desire.

\title{Discovery of a Probable SX Phoenicis Star in M107 (NGC~6171)\altaffilmark{1}}

%% Use \author, \affil, and the \and command to format
%% author and affiliation information.
%% Note that \email has replaced the old \authoremail command
%% from AASTeX v4.0. You can use \email to mark an email address
%% anywhere in the paper, not just in the front matter.
%% As in the title, you can use \\ to force line breaks.

\author{Thayne A. McCombs\altaffilmark{2}}
\affil{Department of Physics and Astronomy, Brigham Young University, Provo,
Utah, 84602}

\author{Erik D. Reinhart\altaffilmark{2}}
\affil{Department of Physics, Willamette University, Salem, Oregon, 97301}
\and

\author{Andrew N. Darragh, Elliott W. Johnson, and Brian W. Murphy}
\affil{Department of Physics and Astronomy, Butler University, Indianapolis,
IN, 46208}

\altaffiltext{1}{Based on observations obtained with the SARA
  Observatory 0.9m telescope at Kitt Peak national Observatory, which is owned
and  operated by the Southeastern Association for Research in Astronomy
  (http://www.saraobservatory.org). }

\altaffiltext{2}{Southeastern Association for Research in Astronomy
(SARA) NSF-REU Summer Intern}
\email{astrothayne@gmail.com, bmurphy@butler.edu}

\begin{abstract}
Using $V$ images taken in May and June 2012 with the SARA Consortium's 0.9-meter telescope located at Kitt Peak National Observatory, we searched for variable stars in the globular cluster M107 (NGC 6171).  The search was accomplished using the ISIS v2.2 image subtraction software.  We refined the positions of the previously known variables and confirmed the 21 RR Lyrae variables
from Clement's Catalog of Variable Stars in Globular Clusters \citep{clement01}.  We also discovered a previously unknown variable which is likely an SX Phoenicis star. For this SX Phoenicis star we measured a fundamental pulsation frequency $19.0122\;\mathrm{day}^{-1}$ ($P=0.05257$ days) and a mean amplitude of $0.046$ magnitudes in the $V$ band.  This variable had an average $V$-band magnitude of 17.72, nearly 2 magnitudes dimmer than the horizontal branch of M107, typical of SX Phoenicis stars and blue stragglers lying just beyond the main sequence turn-off in globular clusters.
\end{abstract}

\keywords{stars: variables: general--Galaxy: globular clusters: individual: NGC 6171}

\section{Introduction}
Given the similar age, distance, and composition of their stars, globular clusters
provide a natural laboratory for studying stellar evolution.  Since globular
clusters are an older population of stars they have a large number of post-main
sequence stars, including many that lie in the instability strip, particularly
RR Lyrae variables. These unstable pulsating variable stars give us insights
into the ways that stars evolve after leaving the main sequence.  In addition by
studying pulsating variables stars and Fourier decomposition of their light
curves it is also possible to determine their composition, mass, and absolute
luminosities \citep{simon93}. 

M107 (NGC 6171) is one such globular cluster containing a number of variable stars. 
Clement's Catalog of Variable Stars in Globular Clusters lists 22 RR Lyrae stars
(15 RR0,7 RR1), a Mira variable, an LB variable, and an unclassified variable in
M107 \citep{clement01}. \citet{clement97} found the periods and light-curve
Fourier coefficients of sixteen of these RR Lyrae stars. Given the relatively
small number of variables previously found in M107 and its location near the
celestial equator where it is observable in both the northern and southern
hemisphere, we chose M107 to search for undiscovered pulsating variable stars.

In previous studies by this group we have searched for variables in 3 other
globular clusters;  identifying, classifying, and producing detailed light
curves of 229 variables with 108 of them being newly discovered
\citep{conroy12,toddy12,darragh12}. Of the total we identified we classified 111 RR0, 79 RR1,
3 RRd/e, 8 SX Phoenicis, 7 eclipsing, and 22 long period variables.   SX
Phoenicis stars are of particular interest.  They are stars similar to a
$\delta$-scuti class variable star but with lower metallicity (Population II),
anomalously large masses and younger ages \citep{rodriguez00}. Compared to RR Lyrae stars
they have a much higher pulsation frequency due to their position near the main
sequence. As mentioned by \citet{mateo90} SX Phe stars are blue stragglers,
likely the results of stellar mergers lying just above the main-sequence turnoff , thus giving past clues to the dynamical history of the cluster.

Using an image subtraction method developed by \citet{alard00} we searched the
central $13'\times13'$ of M107 for variable stars using observations during May
and June 2012. Using this method we confirmed all the previously known variables
in our field, as well as discovered the existence of an SX Phe star in M107.

\section{Observations and Reduction}

Images were obtained using the Southeastern Association for Research in Astronomy (SARA) North 0.9m telescope located at Kitt Peak
National Observatory. The system uses an Apogee Alta U42 CCD detector with
$2048\times2048$ Kodak e2V CC42-40 with a gain of 1.2 electrons per count and a
read noise of 6.3 electrons cooled to approximately -30$^\circ$C. Using
$2\times2$ binning this gave a plate scale of 0.82\arcsec/px and a
$13.6\arcmin \times13.6\arcmin$ field of view. The setup is ideal for finding variables in
globular clusters because the aperture is large enough to find faint variables
but small enough to use reasonable exposure times. In addition we had regular
access to the telescope, and could therefore obtain enough data to analyze the
light curves and determine periods of the variables.

\begin{figure}
    \epsscale{1.25}
    %\plotone{V26_day1.eps}
    %\plotone{V26_day2.eps}
    %\plotone{V26_day3.eps}
    %\plotone{V26_day4.eps}
    \plotone{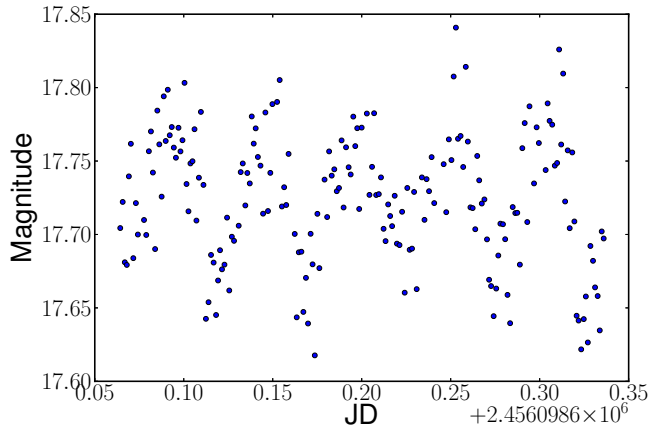}
    \caption{\label{fig:nights}For the night of June 20, 2012, our longest
night,  we plot the measured magnitudes of the SX Phe star against the JD.}
\end{figure}

Observations were made using a Bessel V filter on 23 May 2012 and 7, 8, 9, and
20
June 2012.  Typical seeing was 2.5\arcsec\ and generally between
2\arcsec\ and 3.5\arcsec .  Exposures on 23 May were 120 seconds long, and all exposures
in
June were 90 seconds. Figure \ref{fig:nights} shows the unphased light curve of
a low amplitude SX Phoenicis star on 20 June, our longest night.

We performed standard image reduction using Maxim DL to apply bias, dark, and
flat frames. In order to reduce the number of false positives in ISIS we ran
the frames through Maxim DLs hot and bad pixel removal kernels, and used the
\texttt{craverage} cosmic ray removal task in IRAF. We also rejected image
frames containing satellite trails, star trails from poor guiding, or seeing
worse than 3.5\arcsec.

\section{Analysis}

\subsection{Image Subtraction}

Image subtraction was done using the ISIS v2.2 software package
\citep{alard98,alard00} using the same procedure used by \citet{conroy12}. This
package uses a number of programs and c-shell scripts to align many frames
together, perform image subtraction, find objects with variability, and produce
light curves for those objects. 

ISIS compensates for variable seeing conditions by convolving a high quality
reference frame with each individual frame, and then performing the
subtraction. The {\tt SIGTHRESH} parameter controls the threshold for finding
variable objects. We modified this parameter until roughly a thousand objects
were found, then manually examined the light curves to extract true variables
from the more numerous false positives. 

Our reference frame was produced by combining the 7 best images from the night
of 6 June 2012, because that was the night with the best seeing.

\subsection{Magnitude Calibration}

ISIS outputs its photometry in the form of differential flux (the difference
in flux between the reference frame and the image frame). Unfortunately, as
\citet{baldacci05} pointed out, a simple conversion from the differential flux
to magnitude by 
\begin{equation} 
m=-2.5\log[F_\mathrm{ref}-\Delta F]+C_0
\end{equation} 
does not produce accurate results. In this equation  $F_\mathrm{ref}$ is the
reference flux, $\Delta F$ is the differential flux, and $C_0$ is a constant. To
remedy this we use a method described by \citet{baldacci05} to calibrate the
differential flux from ISIS with instrumental magnitudes from {\tt DAOPHOT}.

The calibration was performed using PSF fitting photometry on the night of 20
June 2012 using the standalone {\tt DAOPHOT}  and {\tt ALLFRAME}
packages\citep{stetson87,stetson94}. In the {\tt DAOPHOT} package we used the
{\tt FIND} procedure to find stars above a threshold of $3.5\sigma$ where
$\sigma$ is the standard deviation of the sky background. We then used the {\tt
PHOT} procedure which performed initial aperture photometry on the stars found.
However since the field is crowded the results of aperture photometry are not
acceptable, therefore we used {\tt PSF} fitting for more accurate photometry.
This was done using {\tt PICK} to select the 25 brightest stars which were not
overexposed and did not overlap with any brighter stars, followed by {\tt PSF}
to construct a semi-empirical model of the PSF for each frame.  {\tt DAOMATCH}
was then used to produce a rough transformation between the frames, followed by
{\tt DAOMASTER} to increase the accuracy of the transformation. After obtaining
transformations 
we created a median image using {\tt MONTAGE2}. Following this step we ran {\tt
FIND}, {\tt PHOT}, {\tt PICK}, and {\tt PSF} on the median image and then {\tt
ALLSTAR} to perform PSF fitting photometry on the median image. Following the
subtraction of the PSFs for all found stars from the original image we again ran
{\tt FIND} and {\tt PHOT} on the subtracted image, and add the new stars to the
star list, and perform {\tt ALLSTAR} on the original image with the revised star
list. This process is repeated until all stars have been added. We then ran {\tt
ALLSTAR} on all of the individual frames. Finally  {\tt ALLFRAME} was applied to
the transformation list to improve the PSF fits for each frame by looking at all
of the frames simultaneously. {\tt DAOPHOT} does not take into account the JD,
airmass, or exposure time, so  scripts were written to extract this information
from the headers.

After obtaining the instrumental magnitudes for the night of 20 June 2012 we
found the instrumental magnitude for the SX Phe star from the reference frame
used in ISIS.  This was accomplished using the procedure outlined above, however
since we only had a single frame we did not perform the steps for {\tt
DAOMATCH}, {\tt DAOMASTER}, {\tt MONTAGE2}, or {\tt ALLFRAME}, and instead used
only {\tt ALLSTAR} to perform the photometry. However multiple runs were used to
find all stars, as was done with the median image. 

Next we attempted to fit the differential flux to the instrumental magnitude for
this night as described by \citet{baldacci05}. Since {\tt DAOPHOT} uses a zero
point of 25 for its photometry we calculated the reference flux for the SX Phe
star from the reference frame using
\begin{equation}
 F=10^{(25-m_\mathrm{inst})/2.5}
\end{equation}
where $m_\mathrm{inst}$ is the instrumental magnitude from the reference frame.
After extracting the instrumental magnitudes from the output of {\tt ALLFRAME},
we used a program written in C++ which attempted several fits of the
differential flux to instrumental magnitude using the equation given by
\citet{baldacci05}:
\begin{equation}
\label{eq:mag-fit}
 m_i=-2.5\log[F_{i,\mathrm{ref}}-A \Delta F_i] + C
\end{equation}
varying A between 0.5 and 2, and C between 24 and 26.  We optimized A and C to
minimize the test statistic
\begin{equation}
 \sum_i\frac{(m_{i,\mathrm{ALLF}}-m_{i,\mathrm{ISIS}})^2}{n}
\end{equation}
where $n$ is the number of acceptable data points. Data points are acceptable if
the residual for the current fit ($m_{i,\mathrm{ALLF}}-m_{i,\mathrm{ISIS}}$) is
less than 0.1 magnitudes. Furthermore any fit where more than 25\% of the data
points are not acceptable is rejected.

Once the values of $A$ and $C$ were obtained, we converted all of the ISIS data
from fluxes to magnitudes using equation \ref{eq:mag-fit}. Although this method
worked
well for the previously discovered variables, the SX Phe was two faint for
{\tt DAOPHOT} to find on the individual frames,
although we could perform photometry for it on the reference frame. Therefore
we used the median values of A and C from the values for the other variables.
This is not a bad estimate since the values for A and C do not vary
significantly in the field of view, and in this case at least, the median, mode
and mean are identical to within 0.01. We obtained the values $A=1.09$ and 
$C=25.19$, both of which are close to the expected values of 1 and 25
respectively.

Finally to transform from instrumental magnitude to apparent magnitude we use
\begin{equation}
 V = V_{inst}-25 + 2.512\log(t_{exp})+22.633%-0.16X
\end{equation}
where $V$ is the apparent magnitude, $V_{inst}$ is the instrumental
magnitude, $t_{exp}$ is the exposure time of the frames used for the
calibration with {\tt DAOPHOT}, which was 90 seconds. The 22.633 term is the
zeropoint of 25 used by {\tt DAOPHOT} minus a measured offset for our system.  This gives us an approximate apparent magnitude of 17.72 for the SX Phe star.

For the RR Lyrae variables that were in our field we applied the same procedure.   We found a median $V$ magnitude for these variables of 15.72, two full magnitudes brighter than our suspected SX Phe star.  Given that RR Lyrae stars are horizontal branch stars this would also be the typical magnitude of the horizontal branch for M107.  Typically the main-sequence turnoff would be 3.5 magnitudes dimmer than the horizontal branch, or roughly 19.2.

\begin{deluxetable}{cc}
 \tabletypesize{\scriptsize}
 \tablecaption{Amplitude variations for the SX Phe star \label{tbl:amps}}
 \tablewidth{\columnwidth}
 \tablehead{
    \colhead{Night of Observation (JD)} &
    \colhead{Amplitude (V Magnitude)}
 }
\startdata
2456070  & 0.054\\
2456085  & 0.052\\
2456086  & 0.053\\
2456087  & 0.032\\
2456098  & 0.046\\
\enddata
\end{deluxetable}

\subsection{Frequency Analysis}

Once our differential fluxes were converted into magnitudes via the procedure
described in section 3.2 they were analyzed using the software package Period04
\citep{period04}. Period04 was used to determine the frequency and amplitude of
the fundamental frequency of the SX Phe star.    We searched for the fundamental
frequency in the range 0 to 50 day$^{-1}$ with a step size of 0.00176
day$^{-1}$.  Two outliers were removed that obviously were spurious points in
the light curve.
The average magnitude of the SX Phe star was also obtained from Period04.   The
results for other harmonics were unreliable due to the faintness of the object
and the quality of the data.

\subsection{Astrometry}

\begin{figure}
  \epsscale{1.18}
 \plotone{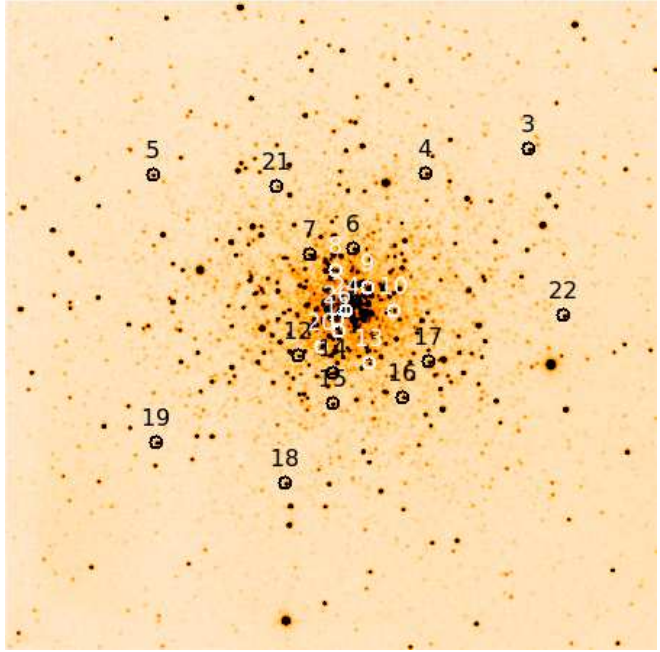}
 \caption{\label{fig:finderall}Positions of variables in M107}
\end{figure}

\begin{figure}
\epsscale{1.18}
 \plotone{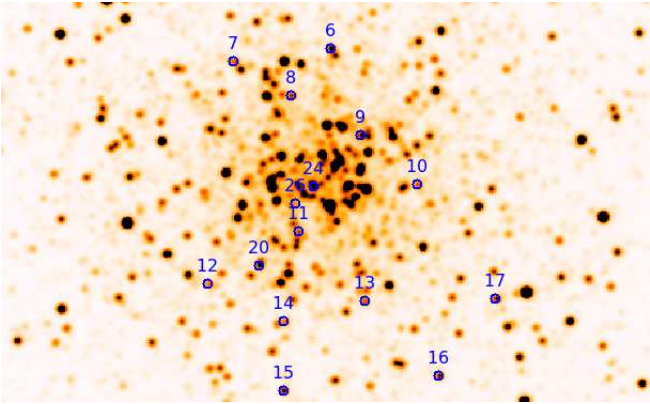}
\caption{\label{fig:core}Positions of variables in the core of M107, star 26 is
the new SX Phe star}
\end{figure}

\begin{figure}

\centering
 \includegraphics[width=\columnwidth]{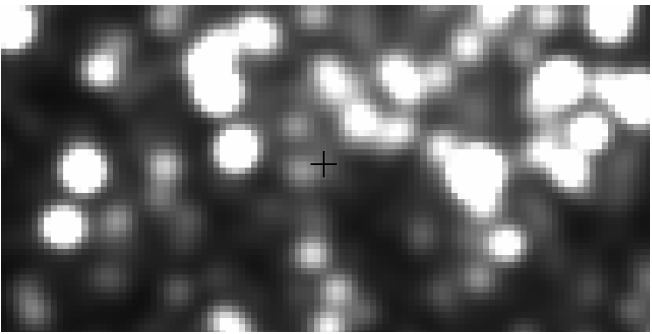}
\caption{\label{fig:v26finder}A finder chart zoomed in on the new SX Phe star
(26)}
\end{figure}

After identifying variable stars using ISIS their coordinates were in units of
pixels.  To convert to conventional celestial coordinates we used the
\texttt{ccmap} and
\texttt{ccsetwcs} tasks in IRAF to apply a world coordinate system to our
alignment image. We used a set of nine stars from the NOMAD catalog to match
the physical coordinates of our frame with the {\tt fk5} celestial coordinate
system. We then used the \texttt{center} task to ensure that the pixel
coordinates for the variable stars were centered on the centroid. Finally we
converted the pixel coordinate system to celestial coordinates. Figures
\ref{fig:finderall}, \ref{fig:core}, and \ref{fig:v26finder} show our field of
view and subsets of our field of view, along with the positions of known
variables. The SX Phe star is indicated as number 26.

\begin{figure}
\epsscale{1.2}
 \plotone{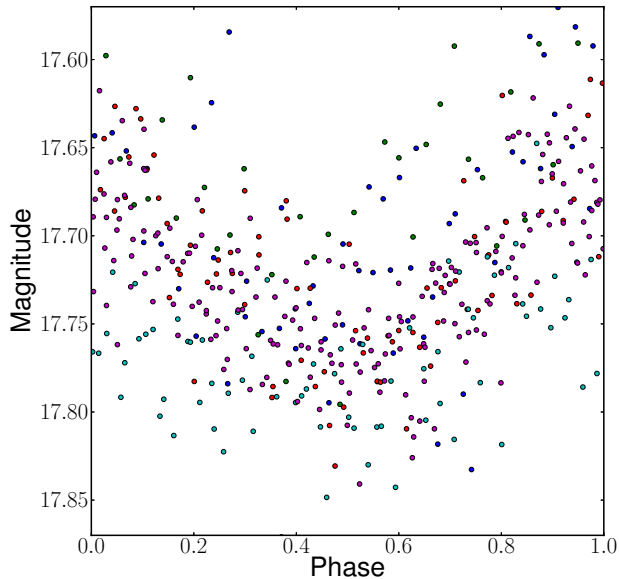}
 \caption{\label{fig:phase} The data from all nights was phased using Period04
with the estimated period of 0.05257 days and plotted together. Different nights
are represented different by colors.}
\end{figure}

\section{Results}
Using the image subtraction package ISIS we found all of the variables documented by
\citet{clement01} in our field and verified their classifications. We also
discovered an SX Phe star located at $16^h32^m32.69^s$, $-13^\circ02'58.92''$.
It was measured to have a pulsation frequency of $19.0122\;\mathrm{day}^{-1}$ (0.05257
days) with an 
amplitude of $0.046$ magnitudes, roughly 0.1 magnitudes from minimum to maximum. The average V magnitude was 17.72. Using
Period04's Monte Carlo simulation we estimated the errors for the frequency and
magnitude as 0.0011 and 0.0031, respectively. The Monte Carlo error for
frequency is less than our exposure time, and the error for magnitude is less
than our error for the transformation, which suggests our results are limited
primarily by observational error.  The frequency, V magnitude, and amplitude are consistent with the variable being an SX Phe star in a globular clusters. It is 2 magnitudes dimmer than the horizontal branch of M107.  This would place it roughly 1.5 magnitudes above the main sequence turn-off which is typically where SX Phe stars and blue stragglers are found. 

Figure \ref{fig:nights} shows our
observations for  June 20 (our longest night) and Figure \ref{fig:phase} shows
all of the data combined and phased.
As can be seen from the Figure \ref{fig:nights} and Figure \ref{fig:phase},
although the frequency appears consistent, the amplitude may be modulating.
Using Period04 we determined amplitudes for each night separately, with the
results listed  in Table \ref{tbl:amps}.

When performing astrometry we discovered that the coordinates
given by \citet{clement01}, which originated from \citet{oosterhoof38} were
inaccurate by a few arcseconds. Given that these coordinates were obtained in
1938 with photographic plates this inaccuracy is not surprising. Here we present
revised coordinates in Table \ref{tbl:astrom}.  Two of the variables, V1 and V2, were outside of our field of view.  Their coordinates are taken from \citet{clement01}.

\begin{deluxetable}{cr@{$^h$}l@{$^m$}l@{$^s$}r@{$^\circ$}l@{$'$}l@{
$''$ } c }
 \tabletypesize{\scriptsize}
\tablecaption{Astrometry and classification of variables in M107
\label{tbl:astrom}}
\tablewidth{\columnwidth}
\tablehead{
    \colhead{Variable\tablenotemark{a}} &
    \multicolumn{3}{c}{RA (2000)\tablenotemark{b}}  &
    \multicolumn{3}{c}{Dec (2000)}  &
    \colhead{Type}
}
\startdata
V1\tablenotemark{c} & 16&32&24.340 & -13&11&55.60 & M \\
V2\tablenotemark{c} & 16&32&41.780 & -13&09&42.40 & RR0 \\ 
 V3 & 16&32&16.656& -13&06&23.60 &  RR0\\
 V4 & 16&32&25.237& -13&05&55.52 &  RR1\\
 V5 &16&32&47.887 & -13&05&57.10 &  RR0\\
 V6 & 16&32&31.290& -13&04&25.60 &  RR1\\
 V7 & 16&32&34.906& -13&04&18.67 &  RR0\\
 V8 & 16&32&32.800& -13&03&59.93 &  RR0\\
 V9 & 16&32&30.181& -13&03&37.84 &  RR1\\
V10 & 16&32&28.054& -13&03&10.32 &  RR0\\
V11 & 16&32&32.574& -13&02&44.98 &  RR0\\
V12 & 16&32&35.984& -13&02&16.57 &  RR0\\
V13 & 16&32&30.095& -13&02&06.09 &  RR0\\
V14 & 16&32&33.118& -13&01&55.53 &  RR0\\
V15 & 16&32&33.182& -13&01&17.45 &  RR1\\
V16 & 16&32&27.326& -13&01&24.86 &  RR0\\
V17 & 16&32&25.146& -13&02&07.07 &  RR0\\
V18 & 16&32&37.163& -12&59&41.75 &  RR0\\
V19 & 16&32&47.839& -13&00&32.52 &  RR1\\
V20 & 16&32&34.060& -13&02&26.23 &  RR0\\
V21 & 16&32&37.579& -13&05&42.22 &  RR1\\
V23 & 16&32&13.917& -13&03&01.46 &  RR1\\
V24 & 16&32&31.958& -13&03&09.73 &  RR0 \\
V25\tablenotemark{d} & \multicolumn{3}{c}{\nodata} & \multicolumn{3}{c}{\nodata}
& Lb \\
V26\tablenotemark{e} & 16&32&32.660& -13&03&00.38 & SX Phe\\
\enddata
\tablenotetext{a}{We used the same variable numbers as \citet{clement01}, with
the exception of V26 which was not in Clement's catalog.}
\tablenotetext{b}{Our coordinates differed by a few arcseconds from
\citet{clement01}, likely due to their using photographic plates.}
\tablenotetext{c}{V1 and V2 are outside of our field of view, the coordinates
and variable type are taken from \citet{clement01}, and the coordinates may be
inaccurate.}
\tablenotetext{d}{V25 was discovered by \citet{lloyd73}, but no coordinates
were available, and we did not detect this star, possibly due to a long period.}
\tablenotetext{e}{V26 was discovered in this study.}
\end{deluxetable}

\section{Conclusion}

We have verified and refined the positions of 21 previously known variables in
M107 as documented by
\citet{clement01}. We have also discovered a probable new SX Phe star located
at 
$16^h32^m32.69^s$, $-13^\circ02'58.92''$ with fundamental frequency of
$19.01221\;\mathrm{day}^{-1}$ (0.05257 days) with a mean V magnitude and 
amplitude of 17.72 and $0.046$ magnitudes, respectively.
The discovery of this faint variable shows the power of using ISIS to look for
variable stars. In addition we have improved the astrometry for the previously
discovered variables in M107.

Unfortunately the SX Phe star was too faint for us to obtain a high
signal-to-noise with our observations, and therefore we were limited in the
amount of frequency analysis we could perform. 

\acknowledgements

We thank C. Alard for making ISIS 2.2 publicly available and P Stetson for useful advise in using DAOPHOT and ALLFRAME. This project
was funded in part by the National Science Foundation Research Experiences for
Undergraduates (REU) program through grant NSF AST-1004872. Additionally A.
Darragh, E. Johnson, and B. Murphy were partially funded by the Butler Institute
for Research and Scholarship. The authors also thank F. Levinson for a generous
gift enabling Butler University's membership in the SARA consortium.

\clearpage

%% The following command ends your manuscript. LaTeX will ignore any text
%% that appears after it.

\end{document}